\documentclass[reprint,amsmath,amssymb,aps,longbibliography]{revtex4-1}
\pdfoutput=1 

\usepackage[T1]{fontenc}
\usepackage[normalem]{ulem}
\usepackage{hyperref}
\usepackage{adjustbox}
\usepackage{multirow}
\usepackage{slashed}
\usepackage{pbox}
\usepackage{makecell}
\usepackage{color}
\usepackage{float}

\begin{document} 

\title{\boldmath Non-minimal Lorentz invariance violation in light of muon anomalous magnetic moment and long-baseline neutrino oscillation data}

\author{Hai-Xing Lin}\email{linhx55@mail2.sysu.edu.cn}
\affiliation{School of Physics, Sun Yat-sen University, Guangzhou 510275, China}

\author{Pedro Pasquini}\email{ppasquini@sjtu.edu.cn}
\affiliation{Tsung-Dao Lee Institute (TDLI), Shanghai Jiao Tong University (SJTU), Shanghai 200240, China}

\author{Jian Tang}\email{tangjian5@mail.sysu.edu.cn}
\affiliation{School of Physics, Sun Yat-sen University, Guangzhou 510275, China}

\author{Sampsa Vihonen}\email{sampsa@mail.sysu.edu.cn}
\affiliation{School of Physics, Sun Yat-sen University, Guangzhou 510275, China}

\date{\today}

\begin{abstract}
    In light of the increasing hints of new physics at the muon $g-2$ and neutrino oscillation experiments, we consider the recently observed tension in the long-baseline neutrino oscillation experiments as a potential indication of Lorentz invariance violation. For this purpose, the latest data from T2K and NO$\nu$A is analysed in presence of non-minimal Lorentz invariance violation. Indeed, we find that isotropic violation in dimensions $D =$ 4, 5 and 6 can alleviate the tension in neutrino oscillation data by about 0.4--2.4$\sigma$ CL significance, with the isotropic coefficient $\gamma^{(5)}_{\tau \tau} =$ 3.58$\times$10$^{-32}$GeV$^{-1}$ yielding the best fit. At the same time, the anomalous muon $g-2$ result can be reproduced with an additional non-isotropic violation of $d^{zt} =$ -1.7$\times$10$^{-25}$. The analysis highlights the possibility of simultaneous relaxation of experimental tensions with Lorentz invariance violation of mixed nature.
\end{abstract}

\maketitle
\flushbottom

\section{\label{sec:intro}Introduction}

Standard Model (SM) of particle physics is the most successful theory to describe properties of elementary particles and their interactions. Its robustness has been tested in numerous experiments, culminating in the discovery of Higgs boson at the LHC. The conservation of Lorentz invariance and {\em CPT} symmetry are an inseparable part of SM physics, as they ensure that physics stay the same regardless of the observer. Only recently, the integrity of SM has started to falter as mounting evidence from electroweak precision observables, CKM matrix element measurements and neutrino experiments are showing divergences between experiments and SM predictions. As physicists look for methods to accommodate SM physics in a more complete theoretical framework, searching evidence of violation to fundamental symmetries could provide hints of the underlying theory, such as the formulation of quantum gravity~\cite{Kostelecky:2009zr,Torri:2020dec,Antonelli:2020nhn}.

{\em CPT} is a fundamental symmetry that is conserved in quantum field theories set in a flat spacetime~\cite{Schwinger:1951xk,Luders:1957bpq}. It is closely related to Lorentz invariance~\cite{Greenberg:2002uu}. Well-known examples of theories that give rise to Lorentz Invariance Violation (LIV) or {\em CPT} violation are found in the string theory~\cite{Kostelecky:1988zi,Kostelecky:1991ak}, while LIV could also arise in supersymmetry together with {\em CPT} violation \cite{Berger:2001rm} and even without it~\cite{Gaete:2021jog}. An example of a simple non-local field theory of {\em CPT} violation can be found in Ref.~\cite{Barenboim:2002tz}. Generally speaking, theories that uphold any non-trivial space-time dependence on the vacuum also lead to violation of either Lorentz invariance, {\em CPT} symmetry, or both. Lorentz invariance can furthermore be violated isotropically or into a specific direction. Effects of LIV and {\em CPT} violation are typically studied in the effective field theory framework, the most general one being the famous Standard Model Extension~\cite{Colladay:1998fq} where SM Hamiltonian is expanded to arbitrary dimensions. Evidence of LIV or CPT-odd operators have been searched in many experiments from atmospheric neutrino fluxes to gravitational waves~\cite{Tomar:2015fha,Stecker:2017gdy,Wei:2018ajw,Rahaman:2021leu,Wang:2021ctl}, which have led to very stringent constraints especially for high-order operators~\cite{Coleman:1998ti,Kostelecky:2008ts}.

In this work, we study the prospects of uncovering LIV physics in neutrino
oscillation experiments. Neutrino oscillations stand out as the first direct
evidence of physics beyond SM, making neutrino experiments an ideal platform to look
for new physics. In recent years, significant advances have been made in the
precision measurements of the standard oscillation parameters: $\theta_{13}$ has
been measured at a few-percent-level at 90\% confidence level (CL) in reactor
experiments, $\theta_{23}$ and $|\Delta m_{31}^2|$ in atmospheric and long-baseline
neutrino experiments, and $\theta_{12}$ and $\Delta m_{21}^2$ in a combination of
solar and reactor neutrino experiments~\cite{Esteban:2020cvm}. At the same time,
observations of neutrino anomalies and tensions in oscillation
data~\cite{LSND:2001aii,Giunti:2010zu,Mention:2011rk,MiniBooNE:2018esg} have seeded
an intense discussion over whether new physics is in
play~\cite{Ohlsson:2012kf,Denton:2020uda,Hu:2020oba,Dasgupta:2021ies}. We investigate the tension in the long-baseline neutrino experiments T2K and NO$\nu$A~\cite{T2K:2021xwb,NOvA:2021nfi}, where recent data has shown
contradicting results on the parameters $\theta_{23}$ and $\delta_{CP}$. We study
the parameter discrepancies in T2K and NO$\nu$A data in light of non-minimal LIV,
focusing on the lesser constrained dimensions 4, 5 and 6. It is shown in this work
that that isotropic LIV could notably alleviate the tension observed in the
three-neutrino mixing. We also show how the recent muon $g-2$
measurements~\cite{Muong-2:2021ojo} could simultaneously arise from the LIV
operators when non-isotropic coefficients are present.

This article is organized as follows: In section\,\ref{sec:theory}, we provide a brief overview of the theoretical formalism of LIV effects in neutrino oscillations and muon $g-2$. In section\,\ref{sec:neutrino_data}, we describe the experimental data and numerical methods adopted in this work. The results on fitting LIV coefficients into T2K and NO$\nu$A data are presented in section\,\ref{sec:results}. We leave the concluding remarks in section\,\ref{sec:conclusions}.


\section{\label{sec:theory}Lorentz Invariance Violation in neutrino oscillations}

In this section, we review the theoretical framework to study LIV in neutrino oscillations. We first consider the isotropic LIV effects in section~\ref{sec:theory:isotropic}. The non-isotropic implementation of LIV is then presented in section~\ref{sec:theory:nonisotropic}. Finally, we discuss the implications to LIV coefficients from recent muon $g-2$ measurements in section~\ref{sec:muon_g-2}.

\subsection{\label{sec:theory:isotropic}Isotropic violation}

In the following, we summarize the
effective Hamiltonian describing neutrino
oscillations under Lorentz invariance
violation. The method described in this section was originally proposed in~\cite{Coleman:1998ti} and further developed in~\cite{Cohen:2006ir,Kostelecky:2011gq,Antonelli:2018fbv}. We focus on LIV terms that appear 
in the kinetic term. The generalized kinetic 
therm of the neutrino Lagrangian 
can be written as
\begin{equation}
\mathcal{L}_{D} 
= 
  i
  \nu^{\dagger}_{iL} 
  \bar\sigma^{\mu} \partial_{\mu}
  \left[
    \delta_{ij} 
  -
    i^{D-4} 
    \gamma_{ij}^{\mu_{1}...\mu_{D-4}}
    \partial_{\mu_{1}}
  ...
    \partial_{\mu_{D-4}}
  \right]
  \nu_{jL} 
  \label{eq:1}
\end{equation}
where $\gamma_{ij}^{\mu_{1}...\mu_{D-4}}$ are tensors of rank $D-4$ that parametrizes the size of the LIV and $\sigma^\mu$ are the Pauli matrices.

In this work, we mainly focus on the isotropic part of the Lorentz violation. The isotropic sector is generally confined by less stringent bounds~\cite{Kostelecky:2008ts}, which makes it more interesting to probe in experiments. Hence, we will set $\mu_i = 0$ unless stated otherwise.
Under the isotropic hypothesis, there is no violation related to direction of the momentum and the neutrino propagation Hamiltonian can therefore be written by two terms, that is,
\begin{eqnarray}
  H 
=
  H_0
+
  H_{\rm LIV},
\end{eqnarray}
where $H_0$ stands for the standard Hamiltonian consistent with SM and $H_{\rm LIV}$ contains the Lorentz-violating terms. In the flavour basis, the standard
Hamiltonian is given by $H_0 = U^\dagger {\rm diag}[0 \Delta m_{21}^2/2E, \Delta m_{31}^2/2E] U + V(x)$ 
and LIV part
by
\begin{equation}
H_{\rm LIV}=
\left(
\begin{array}{ccc}
\gamma^{(D)}_{e e} & \gamma^{(D)}_{e \mu} & \gamma^{(D)}_{e \tau}\\
\gamma^{(D)*}_{e \mu} & \gamma^{(D)}_{\mu \mu}& \gamma^{(D)}_{\mu \tau}\\
\gamma^{(D)*}_{e \tau} & \gamma^{(D)*}_{\mu \tau} & \gamma^{(D)}_{\tau \tau}
\end{array}
\right) E^{D-3}
\label{liv_matrix2}
\,,
\end{equation}
where we simplified our notation to $\gamma_{\alpha \beta}^{0...0} =  \gamma^{(D)}_{\alpha \beta}$.
The parameters $\gamma^{(D)}_{\alpha \beta}$ have mass dimension $4-D$. They are {\em CPT}-odd when $D$ is odd and {\em CPT}-even when $D$ is even. The corresponding Hamiltonian for antineutrinos receives a minus sign when $D$ is an odd number. Parameters $\gamma_{\ell \ell}^{(D)}$ ($\ell = e$, $\mu$, $\tau$) are real and $\gamma_{\ell \ell'}^{(D)}$ ($\ell \neq \ell'$) are complex parameters with phases $\phi_{\ell \ell'}^{(D)}$. In the following, we mainly study LIV parameters in the flavour basis where $\gamma^{(D)}_{\ell \ell'} = |\gamma^{(D)}_{\ell \ell'}| e^{-i\phi_{\ell \ell'}^{(D)}}$.

\subsection{\label{sec:theory:nonisotropic}Non-isotropic violation}

In case of non-isotropic Lorentz invariance violation, the strength of LIV effect depends on the direction of the propagating neutrino. In such case, the equation of state can be decomposed into two parts~\cite{Kostelecky:2003cr}:
\begin{eqnarray}
  \left[ 
    i\delta_{\alpha \beta} \partial_t
    -
      H_0
    -
      \delta H
  \right]
  \nu_\beta
=
  0
  \,,
\end{eqnarray}
where $H_0$ conserves Lorentz invariance and $\delta H$ violates it. Assuming the neutrino propagates into direction $\bf x$, the conserving part can be written as
\begin{eqnarray}
  H_0
=
  - \gamma^0
    \left(
      i \delta_{\alpha \beta}
      \boldsymbol\gamma \cdot \nabla_{\bf x}
    -
      M_{\alpha \beta}
    \right),
\end{eqnarray}
where the neutrino mass matrix is given by $M_{\alpha \beta}$ ($\alpha$, $\beta = e$, $\mu$, $\tau$). The perturbation introducing Lorentz invariance violation on the other hand can be presented as
\begin{equation}
  \delta H 
=
-
  \frac 1 2 
    d_{\alpha \beta}^{\mu 0}
    \left(
        \gamma^0 
        \gamma_5 
        \gamma_\mu
        H_0
    +
        H_0   
        \gamma^0 
        \gamma_5 
        \gamma_\mu
    \right)\\
  -
    i
    d_{\alpha \beta}^{\mu j}
    \gamma^0
    \gamma_5 
    \gamma_\mu
    \partial_j,
\end{equation}
where $d_{\alpha \beta}^{\mu j}$ is the strength of the Lorentz violation into the direction $\vec{x}_j$ ($j =$ 1, 2, 3). We assume neutrino propagation align with direction $\vec{x}_3$ and keep only the axial-current operator related to $\gamma_5\gamma_\mu$. As we shall see in the following section, parameter $d_{\alpha \beta}^{\mu j}$ could have profound implications in the calculation of muon anomalous magnetic moment.

In order to derive the evolution equation for neutrinos, the neutrino mass matrix $M$ must be diagonalized. This can be accomplished within the relativistic limit $|{\bf p}| \gg m_i$, where $p_\mu \approx (|{\bf p}|, - {\bf p})$. In such a case the evolution equation is given by~\footnote{The full derivation of equation\,(\ref{eq:scalarHamiltonian}) can be found in Appendix A of Ref.~\cite{Kostelecky:2003cr}.}
\begin{eqnarray}
  i \partial_t \psi_i 
=
  \left(
  |{\bf p}|
+
  \frac{m_i^2}{2|{\bf p}|}
-
  \frac{1}{|{\bf p}|}
  d^{\mu \nu}_{\alpha \beta} 
  p_\mu p_\nu 
  \right) 
  \psi_i 
\label{eq:scalarHamiltonian}
\end{eqnarray}
where $\psi_i$ is the mass eigenstate of the Hamiltonian $H = H_0 + \delta H$. It is also convenient to express coordinates in the Sun-centered inertial frame~\cite{Kostelecky:2003cr},
\begin{eqnarray}
  i \partial_t \psi_i 
=
  \left(
  |{\bf p}|
+
  \frac{m_i^2}{2|{\bf p}|}
-
  d^{zt}_{ij} 
  |{\bf p}|
  \cos \Theta
  \right) 
  \psi_i,
\label{eq:neutrinoevolution}
\end{eqnarray}
where $\Theta$ denotes colatitude~\footnote{Colatitute is expressed as complement of latitude, that is, $\Phi$ = $\pi - \Theta$.} of the Earth. Neutrino experiments with large variation in colatitude can therefore provide sensitive probes to LIV of this kind.

In neutrino oscillations, we introduce non-isotropic LIV to oscillation probabilities. To do this, we include the non-isotropic coefficient $d^{zt}$ from equation\,(\ref{eq:neutrinoevolution}) in addition to the isotropic LIV parameters $\gamma_{\ell \ell'}^{(D)}$ in the Hamiltonian\,(\ref{liv_matrix2}).

\subsection{\label{sec:muon_g-2}Implications of $(g-2)_\mu$ measurement}

In this subsection, we calculate the contribution to anomalous muon magnetic moment from Lorentz invariance violation. It was recently reported by $g-2$ collaboration that the muon magnetic moment is different from the Standard Model prediction by 4.2$\,\sigma$ CL~\cite{Muong-2:2021ojo}. We show in the following that this value can be accommodated with the non-isotropic LIV coefficient $d_{zt}$ in mass dimension $D =$  5.

The general Lagrangian responsible for LIV in muonic sector can be written as~\cite{Bluhm:1999dx}
\begin{widetext}
\begin{equation}
  \mathcal L_{LIV}
=
  \overline \mu 
  \left[ 
  -
    a_\mu \gamma^\mu 
  +
    b_\mu \gamma_5 \gamma^\mu 
  +
    \frac i 2 
    c^{\alpha \beta} 
    \gamma_\alpha 
    D_\beta
  +
    \frac i 2 
    d^{\alpha \beta} 
    \gamma_5
    \gamma_\alpha 
    D_\beta
  - 
    \frac 1 2
    H^{\alpha \beta}
    \sigma_{\alpha \beta}
  \right]
  \mu 
\label{eq:Muon_LIV_Lagr}
\end{equation}
\end{widetext}
where $a_\alpha$, $b_\alpha$, $c^{\alpha \beta}$, $d^{\alpha \beta}$ and $H^{\alpha \beta}$ parameterize LIV for muons and $D^\alpha = \partial^\mu + A^\mu$ is the covariant derivative. Whereas $a_\alpha$ and $b_\alpha$ are {\em CPT}-odd, other coefficients are {\em CPT}-even. Most of these parameters are strictly constrained by previous experiments~\cite{Kostelecky:2008ts}.

The general contribution to the muon $g-2$ frequency is given by~\cite{Gomes:2014kaa}
\begin{eqnarray}\label{eq:6}
  \delta \omega^{\pm} 
=
  2 \sum_{dnjm} E^{D-3}
  e^{i m \omega_{\oplus} t}
  G_{jm}
  \left[ 
    \check{H}^{(D)}_{nlm}
  \pm  
    \check{g}^{(D)}_{nlm}
  \right],
\end{eqnarray}
where $E^{D-3}$ denotes neutrino energy in dimension $D =$ 3, 4, 5..., $\omega_{\oplus}$ is the Earth's frequency around the Sun and factors $\check{H}^{D}_{nlm}$ and $\check{g}^{D}_{nlm}$ are combinations of parameters $b_\mu$, $d^{\alpha \beta}$ and $H_{\alpha \beta}$. Function $G_{jm} = G_{jm}(\Theta)$ on the other hand depends on colatitude $\Theta$ and vanishes for the isotropic case $j =$ 0. LIV must therefore be non-isotropic to explain the anomalous muon $g-2$ value.

In case of non-isotropic LIV, the contribution to muon $g-2$ arises from mass dimension $D =$ 5. When $j =$ 1, $m =$ 0, the $G_{jm}$ function yields
\begin{eqnarray}\label{eq:7}
  \delta \omega 
=
-
  2 \frac{E^2}{m_\mu}
  G_{10}
  \sqrt{\frac{4\pi}{3}}
  d^{zt}
=
  - 2\frac{E^2}{m_\mu} d^{zt} \cos \Theta,
\end{eqnarray}
where $G_{10}(\Theta) = 0.5\sqrt{3/\pi} \cos\Theta$. The total correction to SM value of muon magnetic moment $a_\mu$ is therefore
\begin{eqnarray}\label{eq:8}
  \Delta a_\mu 
=
  -
    \frac{2E^2\cos \Theta }{eB}
    d^{zt},
\end{eqnarray}
where $B$ is the magnetic field. The $g-2$ collaboration~\cite{Muong-2:2021ojo} reported the value of $\Delta a_\mu = 251\times 10^{-11}$ in their recent measurement, which was acquired in magnetic field $B = 1.45$~T and colatitude $\Theta =$ 48.2$^\circ$~\cite{Keshavarzi:2019bjn}. This correction to muonic $g-2$ can be also obtained with non-isotropic LIV coefficient $d^{zt} =$ -1.7$\times$10$^{-25}$ from equation\,(\ref{eq:8}). Such a coefficient is well within the present experimental bounds from the neutrino sector~\cite{Kostelecky:2008ts}. For more details about the $g-2$ measurement, see also Refs.~\cite{Davier:2010nc,Colangelo:2017qdm,Davier:2017zfy,Davier:2019can,Aoyama:2020ynm}.

\section{\label{sec:neutrino_data}Description of the neutrino oscillation data}

In the present work, we focus on the analysis of the neutrino oscillation data from the presently running Tokai-to-Kamioka (T2K) and NuMI Off-axis $\nu_e$ Appearance (NO$\nu$A) experiments. T2K and NO$\nu$A are long-baseline accelerator based experiments where intensive beams of neutrinos and antineutrinos are created by colliding protons on a fixed target. In this work, we focus on the recent data releases from the T2K and NO$\nu$A collaborations reported in Refs.~\cite{T2K:2021xwb} and~\cite{NOvA:2021nfi}, respectively. A particular area of interest is the observed discrepancy in the $\theta_{23}$ measurement, which has shown a tension in neutrino oscillation data between T2K and NO$\nu$A according to an analysis in the standard neutrino mixing picture. In this section, we briefly review the experimental data and the analysis methods that are used in our work.

\subsection{\label{sec:neutrino_data:T2K}T2K experiment}
Tokai-to-Kamioka (T2K) experiment is one of the two presently running long-baseline neutrino oscillation facilities. T2K uses a proton accelerator of 750~kW average output to generate muon neutrino and antineutrino beams. Based on the J-PARC campus in Tokai, Japan, the neutrino beam traverses 295~km away to Kamioka, where it is met by SuperKamiokande neutrino detector. The neutrino beam is also monitored at the near and intermediate detector facilities ND280 and INGRID, respectively.  Super-Kamiokande and ND280 are both located 2.5$^\circ$ off the beam axis, where the observed neutrinos are mainly of about 600~MeV energy. The beam polarity can be switched between muon neutrino and antineutrino modes, with the initial strategy of dividing the operational time into 2 years in neutrino mode and 6 years in antineutrino mode, respectively. A second run is planned for T2K experiment (T2K-II), with the aim to continue the successful run of the first stage.

The neutrino beam used in T2K consists predominantly of muon neutrinos (96\%-98\%), accompanied by smaller components of beam-related backgrounds. The corresponding antineutrino beam has similar composition with muon antineutrinos taking the majority. Most neutrinos and antineutrinos interact via charged-current quasi-elastic (CCQE) interaction with a small but observable chance for resonant charged-current pion production (CC1$\pi$). The neutrino data collected in SuperKamiokande is typically reported in five different samples: two appearance channels measuring $\nu_\mu \rightarrow \nu_e$ and $\bar{\nu}_\mu \rightarrow \bar{\nu}_e$ oscillations via CCQE interaction and two disappearance channels $\nu_\mu \rightarrow \nu_\mu$ and $\bar{\nu}_\mu \rightarrow \bar{\nu}_\mu$. There is also a third appearance channel dedicated to $\nu_\mu \rightarrow \nu_e$ oscillations observed via $\nu_e$CC1$\pi^+$ interaction. All neutrino events are reconstructed from the information coming from the Cherenkov light when charged particles interact with the water content of the neutrino detectors.

In this work we analyse the neutrino data that has been collected in the first phase of T2K between years 2009 and 2018. In Refs.~\cite{T2K:2021xwb,T2K:2020nqo,T2K:2019bcf}, T2K collaboration reported neutrino oscillation data from 3.13$\times$10$^{21}$ protons-on-target (POT). The collected data contains the information about the reconstructed neutrino and antineutrino events from the combined run of 1.49$\times$10$^{21}$ POT in neutrino beam and 1.64$\times$10$^{21}$ POT in antineutrino beam, respectively. The $\nu_\mu$ and $\bar{\nu}_\mu$ CCQE samples are divided into 28 and 19 energy bins in the interval [0.2, 3.0]~GeV. Correspondingly, $\nu_e$ and $\bar{\nu}_e$ CCQE are scattered over 23 same-size energy bins within [0.1, 1.25]~GeV interval and $\nu_e$CC1$\pi^+$ in 16 bins within [0.45, 1.25]~GeV respectively. All neutrino and antineutrino events were recorded in SuperKamiokande, which is assumed to have 22.5~kton of fiducial mass.

\subsection{\label{sec:neutrino_data:NOvA}NO$\nu$A experiment}

NuMI Off-axis $\nu_e$ Appearance (NO$\nu$A) experiment is the second of the two long-baseline neutrino oscillation experiment currently collecting data. Based in the United States, NO$\nu$A generates beams of muon neutrinos and antineutrinos and sends them to traverse 810~km underground. The neutrino source is based on the NuMI beamline in Fermilab, Illinois, which produces neutrinos with an average beam power of 700~kW. The detector facilities in NOvA are a near detector located 1~km from the beam facility and a far detector stationed at an underground laboratory in Ash River, Minnesota. Both detector facilities are placed 0.8$^\circ$ off-axis from the source. In contrast to T2K, neutrinos and antineutrinos produced in NO$\nu$A spread over a wide range of energies around 2~GeV. Neutrino interactions observed in NO$\nu$A therefore consist of various types of charged-current (CC) interactions. 

The neutrino and antineutrino data analyzed in this work is based on the events that were collected in NO$\nu$A far detector in 2014-2020~\cite{NOvA:2021nfi}. The considered data sets were acquired in the far detector, which is a segmented detector consisting of alternating planes of PVC scintillator. The neutrino detection in NO$\nu$A is based on scintillation light that is emitted by the charged particles created in neutrino-nucleus interactions. The data consist of six different samples which correspond to 12.5$\times$10$^{20}$ POT in neutrino mode and 13.6$\times$10$^{20}$ POT in antineutrino modes. There are four electron-like characterising $\nu_\mu \rightarrow \nu_e$ and $\bar{\nu}_\mu \rightarrow \bar{\nu}_e$ oscillations and two muon-like samples describing $\nu_\mu \rightarrow \nu_\mu$ and $\bar{\nu}_\mu \rightarrow \bar{\nu}_\mu$ disappearance. The electron-like samples are assigned into two categories based on the purity of each event: low-CNN$_{\rm evt}$ and high-CNN$_{\rm evt}$. Here CNN$_{\rm evt}$ stands for the convolutional neural network used in particle identification. The muon-like events on are split into 19 unequally spaced energy bins in the range [0.75, 4.0]~GeV, whereas  the electron-like events are distributed in 6 same-size bins over [1.0, 4.0]~GeV interval. There is also the so-called peripheral sample included in NO$\nu$A, which is used to increase the number of pure electron-like events. In this work, we consider the samples for the electron-like, muon-like and peripheral events while assuming 14~kton fiducial mass in the far detector.

\subsection{\label{sec:neutrino_data:analysis}Numerical analysis}

The analysis of T2K and NO$\nu$A data is based on the $\chi^2$ method. The numerical analysis conducted in this work is done with General Long Baseline Experiment Simulator (GLoBES)~\cite{Huber:2004ka,Huber:2007ji}, which has been modified to calculate neutrino evolution with Lorentz invariance violation. The essential parameters as well as the collected neutrino events are summarized for the T2K and NO$\nu$A experiments in table\,\ref{tab:experiments}.

The neutrino data is analysed with the following $\chi^2$ function:
\begin{equation}
\begin{split}
\chi^2 & = \sum_{i} 2\left[ T_{i,d} - O_{i,d} \left( 1 + \log\frac{O_{i,d}}{T_{i,d}} \right) \right]\\
& + \frac{\zeta_{\text{sg}}^2}{\sigma_{\zeta_{\text{sg}}}^2} + \frac{\zeta_{\text{bg}}^2}{\sigma_{\zeta_{\text{bg}}}^2} + {\rm priors},
\end{split}
\label{LBLChi2}
\end{equation}
where index $i =$ 1, 2, ... runs over the energy bins. Here $O_{i}$ and $T_{i}$ stand for the observed and theoretical (predicted) events in the far detectors of T2K and NO$\nu$A. Nuisance parameters $\zeta_{\text{sg}}$ and $\zeta_{\text{bg}}$ are used in the calculation to reflect systematic uncertainties in the signal and background events, respectively. The systematic uncertainties are addressed with the so-called pull-method~\cite{Fogli:2002pt}. The systematic uncertainties considered in this work influence the predicted events $T_{i}$ in neutrino detectors with a simple shift: $T_{i,d} = (1+\zeta_{\text{sg}})N^{\text{sg}}_{i,d} + (1+\zeta_{\text{bg}})N^{\text{bg}}_{i,d}$, where $N^{\text{sg}}_{i,d}$ and $N^{\text{bg}}_{i,d}$ denote the predicted signal and background events, respectively. The prior function is defined as the Gaussian distributions of each of the standard neutrino oscillation parameters.

Calculation of theoretically predicted events $T_{i}$ is performed entirely by GLoBES. The number of events can be described with the formula
\begin{widetext}
\begin{equation}
    T_i = N_{\rm nucl} \, T \, \epsilon \int_{E_{\rm min}}^{E_{\rm max}}  \int_{E_{\rm min}^{'}}^{E_{\rm max}^{'}} dE \, dE^{'} \, \phi(E) \, \sigma(E) \, R(E, E^{'}) \, P_{\nu_{\ell} \rightarrow \nu_{\ell'}}(E),
\label{eq:LBLevent}
\end{equation}
\end{widetext}
where properties of the neutrino experiment are integrated over the true energy $E$ and reconstructed energy $E'$ of the incident energy. The first part, $N_{\rm nucl} \, T \, \epsilon$, is independent of energy and is defined by the number of nucleons in the neutrino detector, operational time of the experiment and detector efficiency, respectively. The integrand on the other hand is formed by the neutrino flux $\phi(E)$, cross-section $\sigma(E)$, energy resolution function $R(E, E^{'})$ and oscillation probability $P_{\nu_{\ell} \rightarrow \nu_{\ell'}}(E)$. In this work, we adopt the neutrino fluxes and cross-sections for T2K and NO$\nu$A from Refs.~\cite{Cao:2018vdk,T2K:2018rhz} and~\cite{NOvA_fluxes,Messier:1999kj,Paschos:2001np}, respectively.

\begin{table}[!t]
\caption{\label{tab:experiments} Summary of the experimental configurations considered in the numerical analysis. Both experiments describe superbeam configurations of either Water Cherenkov (WC) or Totally Active Scintillator Detector (TASD) technology.}
\begin{center}
\begin{tabular}{c c c}
\hline\hline
\rule{0pt}{3ex}Experiment & T2K & NO$\nu$A  \\ \hline
\rule{0pt}{3ex}Source location & Japan & USA  \\ 
\rule{0pt}{3ex}Beam power & 750~kW & 700~kW \\ 
\rule{0pt}{3ex}Protons-on-target & 3.13$\times$10$^{21}$ & 12.33$\times$10$^{20}$ \\ 
\rule{0pt}{3ex}Detector type & WC & TASD \\ 
\rule{0pt}{3ex}Detector mass & 22.5~kt & 14~kt \\ 
\rule{0pt}{3ex}Baseline length & 295~km & 810~km \\ 
\rule{0pt}{3ex}Off-axis angle & 2.5$^\circ$ & 0.8$^\circ$ \\ 
\rule{0pt}{3ex}Data source & Ref.~\cite{T2K:2021xwb} & Ref.~\cite{NOvA:2021nfi} \\ \hline\hline
\end{tabular}
\end{center}
\end{table}

One of very important elements in the analysis of the neutrino oscillation data is the detector response associated with T2K and NO$\nu$A experiments. In the event calculation, the detector response is mainly represented by function $R(E,E')$ which relates the incident and reconstructed neutrino energies $E$ and $E'$ with a Gaussian function of width $\sigma_{\rm res}(E)$. We analyse the T2K and NO$\nu$A far detector data using a modified energy resolution function, where the Gaussian width is given by $\sigma_{\rm res}(E) = \alpha E + \beta \sqrt{E}$. We furthermore introduce an additional phase shift $\gamma$ in detector response function $R(E,E')$. The energy resolution function is determined for T2K and NO$\nu$A by fitting parameters $\alpha$, $\beta$ and $\gamma$ recursively for each channel until the correct spectral shape is achieved. Remaining inconsistencies between the official data and prediction of GLoBES are mitigated with channel and bin-based efficiencies. 

In order to compute the probabilities $P_{\nu_{\ell} \rightarrow \nu_{\ell'}}$ with general Lorentz invariance violation and $l$, $l' = e$, $\mu$, $\tau$, a custom-made probability code is adopted to include the calculation of isotropic and non-isotropic LIV effects in GLoBES. It should be noted that Lorentz invariance violation of higher dimensions could in principle influence the neutrino fluxes $\phi(E)$. In this work, however, we assume the effect on fluxes to be small and fall within the present uncertainties of the neutrino fluxes.

The systematic uncertainties used in the $\chi^2$ function\,(\ref{LBLChi2}) are one of the key characteristics in the analysis of the neutrino oscillation data. In the analysis of experiment data from T2K experiment, we impose 5\% systematic uncertainty on the signal events undergoing CCQE interaction. The corresponding background events are addressed with 10\% systematic uncertainty. Identical systematic uncertainties are used for the events associated with CC1$\pi$. This choice of priors for the nuisance parameters are found to reproduce the fit results reported by the T2K Collaboration in Ref.~\cite{T2K:2021xwb} with sufficient accuracy. In a similar manner, we implement systematic uncertainties on the each channel considered in the NO$\nu$A experiment. In the analysis of NO$\nu$A far detector data, systematic uncertainties are driven by detector calibration, which amounts to about 5\% systematic uncertainty~\cite{NOvA:2021nfi}. We treat electron-like samples with low and high CNN$_{evt}$ with the same pull parameters. The systematic uncertainties used here are found to be adequate to reproduce the official results reported by the NO$\nu$A Collaboration in Ref.~\cite{NOvA:2021nfi}.

\section{\label{sec:results}Numerical results}

We present the results of our numerical analysis in this section. In this work, we investigated effects of general Lorentz invariance violation (LIV) in dimensions $D =$ 4, 5 and 6. The effects were studied in the context of the $\theta_{23}$ discrepancy recently observed in T2K and NO$\nu$A. In the following, we examine whether LIV could alleviate the observed tension between the recent data in T2K and NO$\nu$A and improve the fit to the standard neutrino oscillation parameters. At the same time, we study the effect of LIV in the anomalous muon magnetic moment. As we shall see in this section, isotropic LIV with dimension $D =$ 5 provides the best fit result to the T2K and NO$\nu$A data while its non-isotropic version is simultaneously able to resolve $(g-2)_\mu$.

We first begin investigation with the isotropic Lorentz invariance violation. The analysis of the T2K and NO$\nu$A far detector data is carried out in dimensions $D =$ 4, 5 and 6. The first goal of this study is to identify the LIV parameters that lead to significant improvement on the goodness-of-fit in the T2K and NO$\nu$A data. The second goal is to determine the resulting effect on the fit values of $\sin^2 \theta_{23}$ and $\delta_{CP}$.

\begin{widetext}

\begin{table}[!ht]
\caption{\label{tab:fit_results}Fit results from the scan of isotropic Lorentz invariance violation in dimension $D =$ 4, 5 and 6. The results are obtained from a combined fit to T2K and NO$\nu$A far detector data. The consequent improvement to the fit is shown in the $\Delta \chi^2$ and significance columns. Results are obtained assuming normal ordering for neutrino masses.}
\begin{center}
\begin{tabular}{cccccc}\hline\hline
Parameter & Fit value ($\sin^2 \theta_{23}$) & Fit value ($\delta_{CP}/\pi$) & Fit value ($|\gamma_{\ell \ell'}|$) & $\Delta \chi^2$ & Significance (C.L.)\\ \hline
\rule{0pt}{3ex}$\gamma_{e e}^{(4)}$ & 0.562 & 1.33 & -4.62$\times$10$^{-23}$ & 0.89 & 0.95$\sigma$ \\ 
\rule{0pt}{3ex}$\gamma_{e \mu}^{(4)}$ & 0.562 & 1.48 & 0.88$\times$10$^{-23}$ & 3.96 & 1.48$\sigma$ \\ 
\rule{0pt}{3ex}$\gamma_{e \tau}^{(4)}$ & 0.563 & 1.49 & 0.95$\times$10$^{-23}$ & 4.24 & 1.56$\sigma$ \\ 
\rule{0pt}{3ex}$\gamma_{\mu \mu}^{(4)}$ & 0.522 & 1.30 & 5.58$\times$10$^{-23}$ & 2.47 & 1.57$\sigma$ \\ 
\rule{0pt}{3ex}$\gamma_{\mu \tau}^{(4)}$ & 0.549 & 1.43 & -3.75$\times$10$^{-23}$ & 4.80 & 1.69$\sigma$ \\ 
\rule{0pt}{3ex}$\gamma_{\tau \tau}^{(4)}$ & 0.543 & 1.22 & -2.65$\times$10$^{-23}$ & 0.57 & 0.75$\sigma$\\ \hline
\rule{0pt}{3ex}$\gamma_{e e}^{(5)}$ & 0.561 & 1.48 & -4.65$\times$10$^{-32}$\,GeV$^{-1}$ & 4.34 & 2.08$\sigma$ \\ 
\rule{0pt}{3ex}$\gamma_{e \mu}^{(5)}$ & 0.561 & 1.50 & 0.28$\times$10$^{-32}$\,GeV$^{-1}$ & 3.85 & 1.45$\sigma$ \\ 
\rule{0pt}{3ex}$\gamma_{e \tau}^{(5)}$ & 0.559 & 1.49 & 1.17$\times$10$^{-32}$\,GeV$^{-1}$ & 4.41 & 1.60$\sigma$ \\ 
\rule{0pt}{3ex}$\gamma_{\mu \mu}^{(5)}$ & 0.524 & 1.31 & -3.32$\times$10$^{-32}$\,GeV$^{-1}$ & 2.73 & 1.65$\sigma$ \\ 
\rule{0pt}{3ex}$\gamma_{\mu \tau}^{(5)}$ & 0.560 & 1.26 & -0.45$\times$10$^{-32}$\,GeV$^{-1}$ & 0.65 & 0.36$\sigma$ \\ 
\rule{0pt}{3ex}$\gamma_{\tau \tau}^{(5)}$ & 0.526 & 1.45 & 3.58$\times$10$^{-32}$\,GeV$^{-1}$ & 5.79 & 2.41$\sigma$ \\ \hline
\rule{0pt}{3ex}$\gamma_{e e}^{(6)}$ & 0.562 & 1.34 & -1.15$\times$10$^{-41}$\,GeV$^{-2}$ & 1.00 & 1.00$\sigma$ \\ 
\rule{0pt}{3ex}$\gamma_{e \mu}^{(6)}$ & 0.560 & 1.43 & 0.11$\times$10$^{-41}$\,GeV$^{-2}$ & 3.05 & 1.23$\sigma$ \\ 
\rule{0pt}{3ex}$\gamma_{e \tau}^{(6)}$ & 0.560 & 1.46 & 0.31$\times$10$^{-41}$\,GeV$^{-2}$ & 4.90 & 1.72$\sigma$ \\ 
\rule{0pt}{3ex}$\gamma_{\mu \mu}^{(6)}$ & 0.548 & 1.25 & 0.81$\times$10$^{-41}$\,GeV$^{-2}$ & 1.32 & 1.15$\sigma$ \\ 
\rule{0pt}{3ex}$\gamma_{\mu \tau}^{(6)}$ & 0.554 & 1.27 & -0.17$\times$10$^{-41}$\,GeV$^{-2}$ & 1.29 & 0.64$\sigma$ \\ 
\rule{0pt}{3ex}$\gamma_{\tau \tau}^{(6)}$ & 0.544 & 1.22 & -0.28$\times$10$^{-41}$\,GeV$^{-2}$ & 0.15 & 0.39$\sigma$ \\ \hline\hline
\end{tabular}
\end{center}
\end{table}

\end{widetext}

The analysis of the T2K and NO$\nu$A data is carried out as follows. Using the methods described in section\,\ref{sec:neutrino_data}, the far detector data in the two experiments is fitted with a $\chi^2$ function. We fix the solar parameters $\theta_{12}$ and $\Delta m_{21}^2$ to their respective best-fit values 33.4$^\circ$ and 7.4$\times$10$^{-5}$eV$^2$ following the most recent global fit~\cite{Esteban:2020cvm}. We also impose a prior from the reactor experiments: $\sin^2 2\theta_{13} =$ 0.0857$\pm$0.0046. The $\chi^2$ distribution is computed for parameters $\gamma^{(D)}_{\ell \ell'} = |\gamma^{(D)}_{\ell \ell'}| \exp{(-i \phi^{(D)}_{\ell \ell'})}$ keeping one LIV parameter free each time. To see the difference to the standard fit, we also obtain the $\chi^2$ distribution corresponding to the case where all LIV parameters are kept at zero.

The fit results to $\sin^2 \theta_{23}$ and $\delta_{CP}$ from the joint analysis is T2K and NO$\nu$A data are presented in table\,\ref{tab:fit_results}. The LIV parameters $\gamma_{\ell \ell'}^{(D)}$ were analysed one by one in dimensions $D =$ 4, 5 and 6. In each row, the denoted parameter was let to run free in order to obtain the fit result. The resulting improvement to the fit result is listed in the $\Delta \chi^2$ column, where $\Delta \chi^2 = \chi^2_{SM} - \chi^2_{LIV}$ is calculated from $\chi^2$ values attributed to the standard and LIV-influenced fits, respectively. Finally, the significance column indicates the confidence level (C.L.) at which the fit obtained in presence of the LIV parameter is favoured over the standard fit result. The best fit result is obtained with parameter $\gamma^{(5)}_{\tau \tau}$, which can yield an enhancement of about 2.4$\,\sigma$ C.L. in statistical significance. All results are consistent with the experimental bounds reported in the literature~\cite{Kostelecky:2008ts}, including the stringent bounds from IceCube~\cite{IceCube:2017qyp}. The statistical significance is computed using Wilks' theorem~\cite{Wilks:1938dza}. For the caveats concerning the statistical analysis in T2K and NO$\nu$A data, see Ref.~\cite{Algeri:2019lah}.

The effects of the Lorentz invariance violation on the $\theta_{23}$ and $\delta_{CP}$ measurements in T2K and NO$\nu$A experiments are shown in Fig.~\ref{fig:discr_evol}. The standard fits on T2K and NO$\nu$A data are illustrated with solid red and blue lines, respectively, whereas the fits obtained with Lorentz invariance violation are indicated with dashed lines. The discrepancy in the $\theta_{23}$ measurements is evident in the standard fits, which are clearly separated in T2K and NO$\nu$A. This tension is partially removed with the introduction of LIV coefficient $\gamma^{(5)}_{\tau \tau}$. Imposing LIV into the fit leaves the T2K fit mostly unaffected but shifts the NO$\nu$A fit significantly towards the $\theta_{23}$ value that is preferred by the T2K data. This results in an alleviation of tension in the T2K and NO$\nu$A measurements. The tension between the T2K and NO$\nu$A data can also be seen the fit results on $\delta_{CP}$, which show nearly $\Delta \chi^2 \simeq$ 5 difference at the value that is currently preferred by T2K data. Marginalizing over $\gamma^{(D)}_{\tau \tau}$ reduces this difference to $\Delta \chi^2 \simeq$ 2. This amounts to about 1.8$\,\sigma$ C.L. improvement.
\begin{widetext}
\centering
\begin{minipage}{\linewidth}
\begin{figure}[H]
        \center{\includegraphics[width=\textwidth]{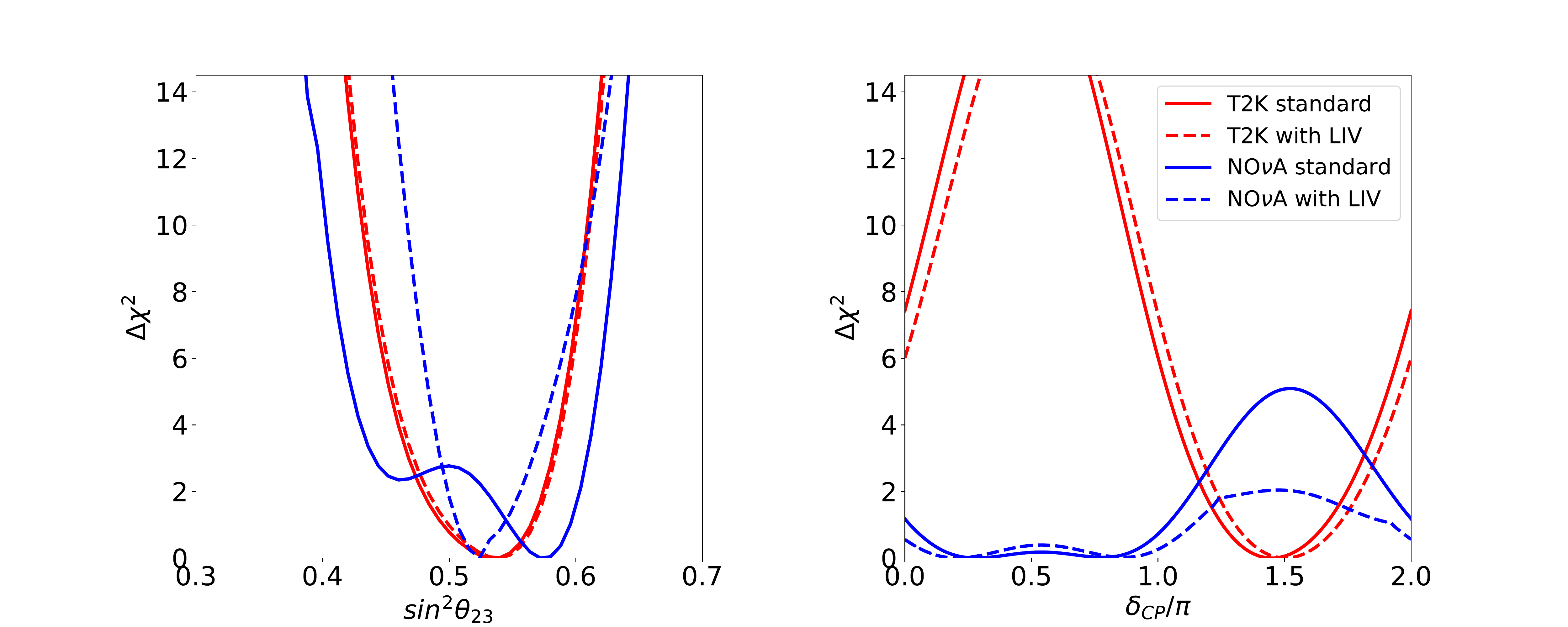}}
        \caption{The effect of Lorentz invariance violation on the determination of $\theta_{23}$ (left panel) and $\delta_{CP}$ (right panel) in T2K and NO$\nu$A experiments. Fit results are shown for T2K and NO$\nu$A experiments with red and blue lines, respectively. Solid lines correspond to the standard fit whereas dashed lines represent fit results where $\gamma^{(5)}_{\tau \tau}$ is taken as a free parameter.}
        \label{fig:discr_evol}
\end{figure}
\end{minipage}
\end{widetext}
\noindent 

\begin{widetext}

\begin{figure}[H]
        \center{\includegraphics[width=\textwidth]{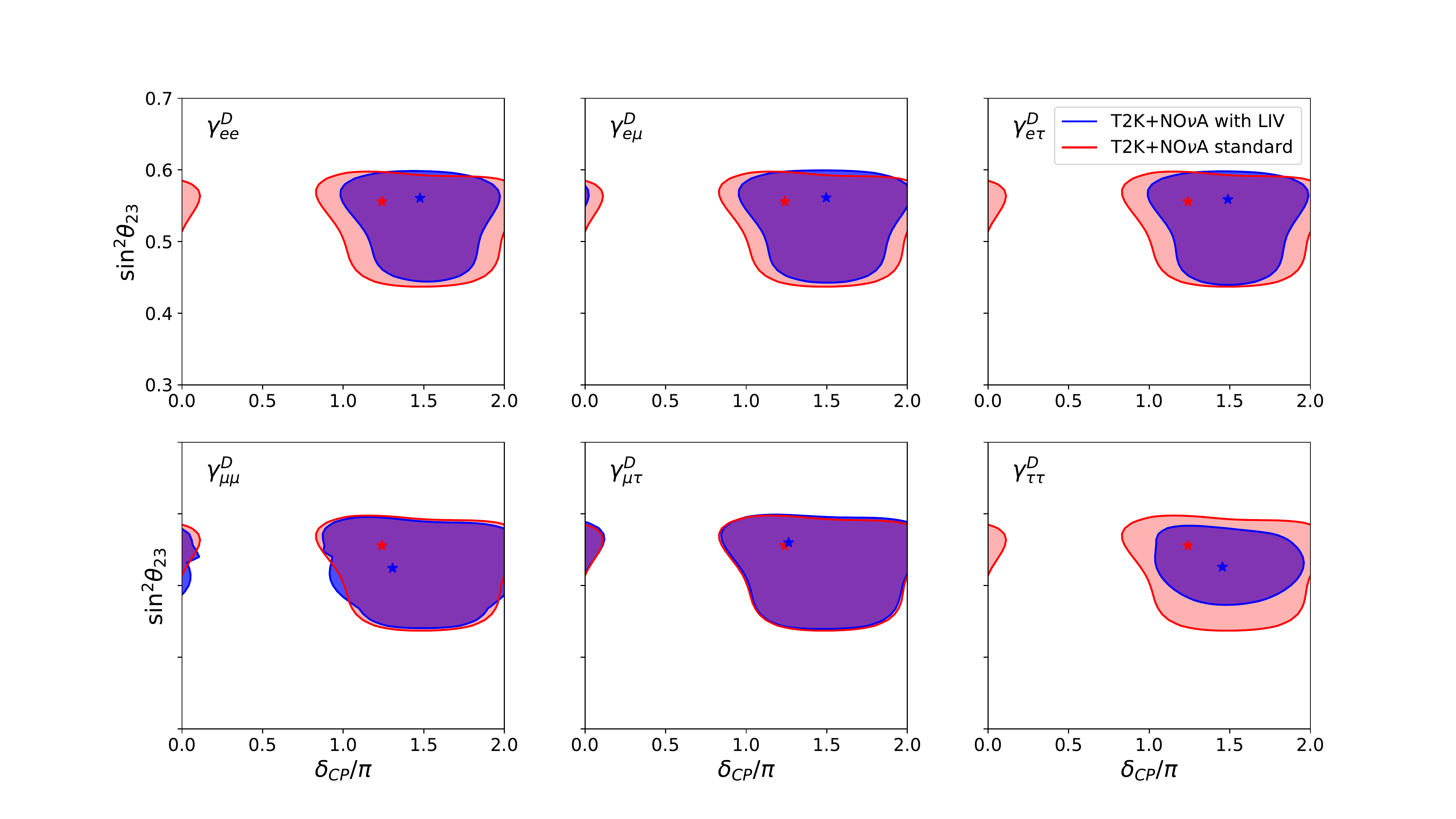}}
        \caption{Allowed values on the $\delta_{CP}-\sin^2\theta_{23}$ plane according to the neutrino oscillation data in T2K and NO$\nu$A experiments. Isotropic Lorentz invariance violation of dimension $D = 5$ is included in the parameter region highlighted with blue colour, whereas standard oscillations are shown with red colour. The best-fit values identified in the two scenarios are indicated with blue and red stars, respectively. The results are shown by 90\% C.L. while assuming normally ordered neutrino masses.}
        \label{fig:isotr-dim5}
\end{figure}

\end{widetext}

The impact of each LIV parameter on the T2K and NO$\nu$A analysis is further elaborated in Fig.~\ref{fig:isotr-dim5}. Each panel portrays the fit to $\delta_{CP} - \sin^2\theta_{23}$ plane as obtained from the joint analysis of the two experiments. The red curves represent the edges of the enclosed parameter space in the standard neutrino oscillations, as allowed by 90\% C.L. in T2K and NO$\nu$A. The best-fit point is marked with the red dot. On the other hand, the corresponding limits obtained with the freely-varying LIV parameters are presented with blue colour. The results show how the best-fit points shift upon the introduction of the single LIV coefficient for $D =$ 5. The fits reveal that the best fit values for $\sin^2\theta_{23}$ and $\delta_{CP}$ shift significantly for parameters $\gamma^{(D)}_{\mu \mu}$ and $\gamma^{(D)}_{\tau \tau}$, whereas the inclusion of $\gamma^{D}_{e e}$, $\gamma^{D}_{e \mu}$ or $\gamma^{D}_{e \tau}$ only has notable effect on the fit value of $\delta_{CP}$. The last off-diagonal parameter $\gamma^{D}_{\mu \tau}$ on the other hand leads to no significant changes in the fit values of $\sin^2\theta_{23}$ and $\delta_{CP}$. The fit results obtained for dimensions $D =$ 4 and 6 are analogous and are not shown here.

We finally make a remark on the effect of non-isotropic Lorentz invariance violation in neutrino experiments. As we pointed out in section\,\ref{sec:theory:nonisotropic}, the recently measured anomalous muon magnetic moment~\cite{Muong-2:2021ojo} could be explained with non-isotropic LIV coefficient $d^{zt} = -$1.7$\times$10$^{-25}$. We investigated the potential effect of this solution in the long-baseline experiments T2K and NO$\nu$A. Our results show non-isotropic coefficient too small to induce notable effects in T2K and NO$\nu$A~\footnote{The smallness of the non-isotropic effect is mainly owed to the relatively small difference in colatitudes between T2K and NO$\nu$A. In T2K, colatitude is given by $\cos \Theta =$ 0.993 whereas in NO$\nu$A it is $\cos \Theta =$ 1.000. Future experiments with longer baseline lengths could see more profound effect from non-isotropic Lorentz invariance violation.}, where the value of $d^{zt}$ required to satisfy $(g-2)_\mu$ leads only to $\chi^2_{\rm min} \sim$ 10$^{-3}$ correction to the fit result. It is therefore possible to recover the measured $(g-2)_\mu$ value with non-isotropic LIV and alleviate the tension on $\theta_{23}$ in T2K and NO$\nu$A at the same time.

\section{\label{sec:conclusions}Conclusions}

Lorentz invariance violation could have profound effects in the interpretation of physical observations in laboratories and astrophysical environments. In the present work, we have investigated non-minimal LIV as a potential solution to the recently observed tension in the measurement of the atmospheric mixing angle $\theta_{23}$ and Dirac {\em CP} phase $\delta_{CP}$ in the long-baseline neutrino experiments T2K and NO$\nu$A. To this extent, we interpreted the recently published experiment data in terms of isotropic and non-isotropic LIV effects.

In contrast to the previous studies conducted on the topic, we studied the relatively unexplored Lorentz invariance violation in dimensions $D =$ 4, 5 and 6. Investigating the isotropic effect on the fits to T2K and NO$\nu$A data, we found that diagonal parameters $\gamma^{(D)}_{\ell \ell'}$ ($\ell$, $\ell' = e$, $\mu$ and $\tau$) could resolve the tension by about 0.4--2.4$\,\sigma$ confidence levels when one parameter is considered at a time. The extracted fit results are consistent the existing bounds on Lorentz invariance violation in the neutrino sector~\cite{Kostelecky:2008ts}. We found the best fit with $\gamma^{(5)}_{\tau \tau} = $ 3.58$\times$10$^{-32}$GeV$^{-1}$, which leads to an improvement of 2.41$\,\sigma$ C.L. from the standard scenario where Lorentz invariance remains conserved. In this case, the favoured values for the atmospheric mixing and Dirac {\em CP} phase are $\sin^2\theta_{23} =$ 0.526 and $\delta_{CP} =$ 1.45$\pi$, respectively.

Lorentz invariance violation is also able to explain the recent results on the anomalous muon magnetic moment as reported by the Muon $g-2$ Collaboration~\cite{Muong-2:2021ojo}. The measured value of $(g-2)_\mu$ could be generated by non-isotropic Lorentz invariance violation with directional coefficient $d^{zt} \simeq -$1.7$\times$10$^{-25}$. Isotropic Lorentz invariance violation on the other hand has no effect on the muon magnetic moment. We estimated the impact of this specific non-isotropic Lorentz invariance violation on the neutrino oscillations in T2K and NO$\nu$A. We find the effect of the non-isotropic LIV parameter to be indistinguishable in T2K and NO$\nu$A due to relatively small change in colatitude in their experimental setups. Though neutrino oscillations and muon $g-2$ do not favour the same type of LIV, it is noteworthy that both solutions can exist simultaneously.

In summary, we have shown that non-minimal Lorentz invariance violation can notably alleviate the $\theta_{23}$ discrepancy in T2K and NO$\nu$A data and simultaneously give rise to the anomalous muon magnetic moment. Respecting the present experimental bounds, we note that the significance for isotropic LIV can be as large as 2.41$\sigma$ C.L. {Our results place a mild preference on dimension-5 LIV. We are looking forward to the future long-baseline neutrino oscillation data to check whether the observed discrepancy is due to the violation of Lorentz invariance.

\vspace{0.5cm}

\acknowledgments
This project was supported in part by National Natural Science Foundation of China under Grant Nos. 12075326, 12090064, 11505301 and 11881240247. PP is additionally supported by the Shanghai Pujian Program under Grant No. 20PJ1407800 and SV by China Postdoctoral Science Foundation under Grant No. 2020M672930. JT and PP appreciate the support from CAS Center for Excellence in Particle Physics (CCEPP).

\appendix 

\section{\label{app:inverted}Implications on mass ordering}

In the present work, we have conducted the analysis on T2K and NO$\nu$A data assuming normal ordering for neutrino masses. For completeness, we now consider the results in the case of inverted ordering and discuss the implications on the sensitivity to neutrino mass ordering.

Figure\,\ref{fig:inverted} provides the fit results on the $\delta_{CP} - \sin^2 \theta_{23}$ plane at 90\% C.L. in T2K and NO$\nu$A experiments when
\begin{widetext}
\centering
\begin{minipage}{\linewidth}
\begin{figure}[H]
        \center{\includegraphics[width=\textwidth]{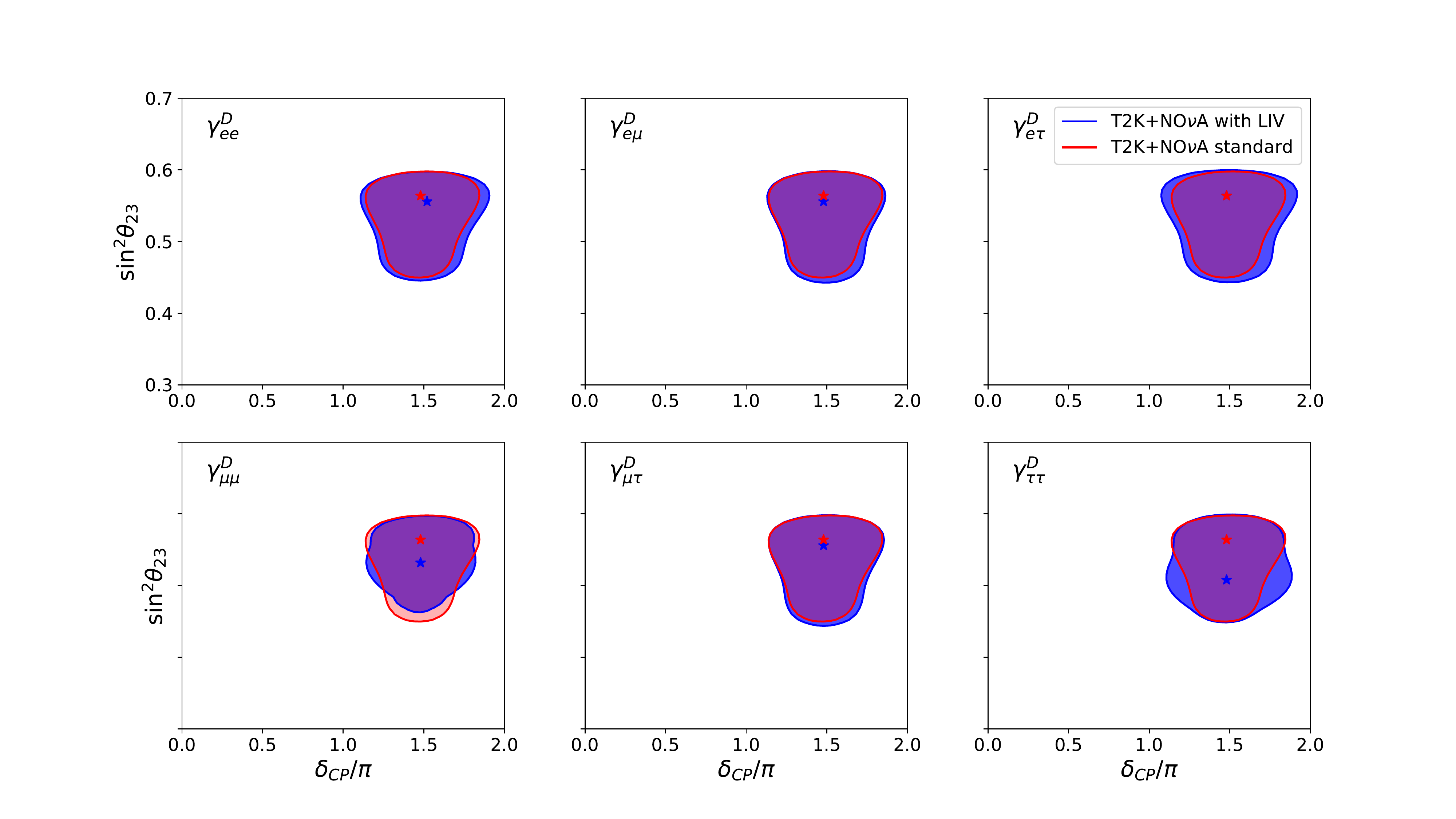}}
        \caption{Fit results on the $\delta_{CP}-\sin^2\theta_{23}$ plane when inverted ordering is assumed. Neutrino data from the T2K and NO$\nu$A experiments are analysed in the standard three-neutrino picture (red colour) and also letting $\gamma^{(D)}_{\ell \ell'}$ run free, whereby $D =$ 5 and $\ell$, $\ell' = e$, $\mu$ and $\tau$. Results are shown at 90\% C.L. while the best-ft values are marked by stars.}
        \label{fig:inverted}
\end{figure}
\end{minipage}
\end{widetext}
\noindent inverted mass ordering is assumed. The standard three-neutrino oscillation scenario is taken into account in the red regions, which correspond to the scenario where no Lorentz invariance violation takes place. Isotropic LIV is taken into account in the blue regions, which were obtained by letting $\gamma^{(D)}_{\ell \ell'}$ run free for $D =$ 5 and $\ell$, $\ell' = e$, $\mu$ and $\tau$. Each panel corresponds to the running of one
\begin{widetext}
\centering
\begin{minipage}{\linewidth}
\begin{figure}[H]
        \center{\includegraphics[width=0.88\textwidth]{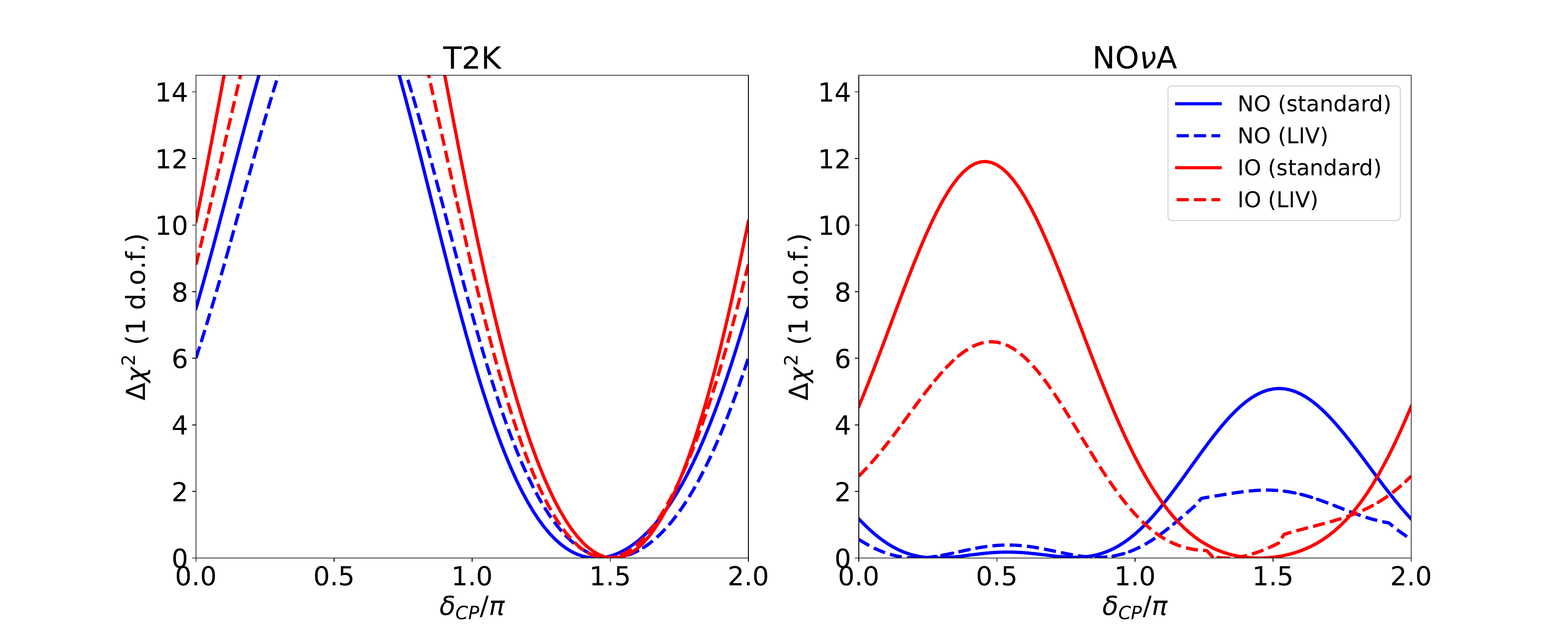}}
        \caption{The effect of Lorentz invariance violation on the determination of mass hierarchy in T2K and NO$\nu$A. The fits to the far detector data in T2K (left panel) and NO$\nu$A (right panel) are presented for normal ordering (NO) and inverted ordering (IO). In the standard scenario (solid) no violation is taken into account, while $\gamma_{\tau \tau}^{(5)}$ is marginalized in the LIV scenario (dashed).}
        \label{fig:mass_ordering}
\end{figure}
\end{minipage}
\end{widetext}
\noindent LIV parameter, whilst other LIV parameters are fixed at zero. We have similarly performed the analysis for LIV parameters in dimensions $D =$ 4 and 6, finding analogous results.

The effects of LIV is studied in the sensitivity to neutrino mass ordering by comparing the fit results in both neutrino mass orderings. The results are presented in Figure\,\ref{fig:mass_ordering}, where the fit results are shown separately for T2K and NO$\nu$A in left and right panels, respectively. While the fit results associated with normal ordering (NO) are shown in blue colour, the results corresponding to inverted ordering (IO) are indicated with red colour. The sensitivity to the neutrino mass ordering can be seen from the relative difference of $\Delta \chi^2$ between the NO and IO curves. In the standard oscillation scenario (solid curves), the difference between NO and IO is more significant in NO$\nu$A than in T2K. The higher sensitivity in NO$\nu$A arises mainly from the longer baseline length in the experiment. When the LIV effect is taken into account (dashed curves), the LIV parameter $\gamma_{\tau \tau}^{(5)}$ is let to vary freely. In such case, the difference between NO and IO is nearly halved for most values of $\delta_{CP}$, indicating significant loss in mass hierarchy discrimination.

\bibliography{references.bib}

\end{document}